\documentclass{article}
\usepackage{ijcai22}

\usepackage{times}

\usepackage{soul}
\usepackage{url}
\usepackage[hidelinks]{hyperref}
\usepackage[utf8]{inputenc}
\usepackage[small]{caption}
\usepackage{graphicx}
\usepackage{amsmath}
\usepackage{booktabs}
\usepackage{algorithm}
\usepackage{algpseudocode}
\usepackage[frozencache,cachedir=.]{minted}
\usepackage{comment}
\usepackage{xspace}
\usepackage{xcolor}
\usepackage[framemethod=tikz]{mdframed}

\usepackage{subcaption}

\newtheorem{definition}{Definition}

\algnewcommand\algorithmicforeach{\textbf{for each}}
\algdef{S}[FOR]{ForEach}[1]{\algorithmicforeach\ #1\ \algorithmicdo}

\title{On Repairing Natural Language to SQL Queries}

\author{
Aidan Z.H. Yang$^3$\footnote{Contact Author}\and
Ricardo Brancas$^1$\And
Pedro Esteves$^{1}$\And
Sofia Aparicio$^2$\And
Joao Pedro Nadkarni$^2$\And
Miguel Terra-Neves$^2$\And
Vasco Manquinho$^1$\And
Ruben Martins$^3$
\affiliations
$^1$ INESC-ID, IST - Universidade de Lisboa, Portugal\\
$^2$ OutSystems, Portugal\\
$^3$ Carnegie Mellon University, USA\\
\emails
\{ricardo.brancas,pedro.f.esteves,vasco.manquinho\}@tecnico.ulisboa.pt$^1$,
\{sofia.aparicio,joao.nadkarni,miguel.neves\}@outsystems.com$^2$,
\{aidanyan,rubenm\}@andrew.cmu.edu$^3$
}

\newcommand{\ratsql}{RAT-SQL\xspace}
\newcommand{\smbop}{SmBoP\xspace}
\newcommand{\patsql}{\textsc{PatSQL}\xspace}

\mdfdefinestyle{mpdframe}{
    frametitlebackgroundcolor   =black!15,
    frametitlerule              =true,
    roundcorner                 =5pt,
    middlelinewidth             =1pt,
    innermargin                 =0.3cm,
    outermargin                 =0.3cm,
    innerleftmargin             =0.3cm,
    innerrightmargin            =0.3cm,
    innertopmargin              =0.3cm,
    innerbottommargin           =0.3cm
}

\begin{document}

\maketitle

\begin{abstract}
Data analysts use SQL queries to access and manipulate data on their databases. However, these queries are often challenging to write, and small mistakes can lead to unexpected data output. Recent work has explored several ways to automatically synthesize queries based on a user-provided specification. One promising technique called text-to-SQL consists of the user providing a natural language description of the intended behavior and the database's schema. Even though text-to-SQL tools are becoming more accurate, there are still many instances where they fail to produce the correct query.

In this paper, we analyze when text-to-SQL tools fail to return the correct query and show that it is often the case that the returned query is close to a correct query. We propose to repair these failing queries using a mutation-based approach that is agnostic to the text-to-SQL tool being used. We evaluate our approach on two recent text-to-SQL tools, \ratsql and \smbop, and show that our approach can repair a significant number of failing queries.

\end{abstract}

\section{Introduction}
\label{sec:intro}

Data analysts are usually domain experts, but they often lack the programming skills
to write their own code for data manipulation or to query databases. Hence, several low-code
platforms for software development have been built that use high-level
interfaces for users to specify their intent.
In these platforms, user intent can be specified by either a natural
language description~\cite{sqlizer,DBLP:conf/acl/WangSLPR20,DBLP:conf/naacl/RubinB21} or input-output examples~\cite{scythe,orvalhoVLDB20,patsql}.
%among other~\cite{sygus,mars,synquid}.

In recent years, several new approaches have been proposed to improve text-to-SQL tools~\cite{DBLP:conf/acl/WangSLPR20,lin2020bridging,DBLP:conf/naacl/RubinB21,DBLP:conf/emnlp/ScholakSB21,DBLP:conf/emnlp/GanCXPWDZ21,DBLP:conf/acl/CaoC0ZZ020,DBLP:conf/aaai/ShiNWZLWSX21}. The user gives as input a database and a description in natural language of the desired task. The text-to-SQL tool is able to return a query that matches the user's intent. For instance, consider that a user has a database of airlines and gives the text-to-SQL tool the following natural language description: ``\emph{Find the name of the airline which runs the most number of routes}''. Ideally, a text-to-SQL returns the following SQL query:

\begin{minted}{sql}
SELECT airlines.name, routes.alid 
FROM routes JOIN airlines 
    ON routes.alid = airlines.alid
GROUP BY airlines.name
ORDER BY Count(*) DESC
LIMIT 1
\end{minted}

Text-to-SQL tools can solve problems of this kind of difficulty and have an accuracy of over 70\% on public datasets.\footnote{\url{https://yale-lily.github.io/spider}}
However, despite its recent improvements, these tools still fail to provide the correct query in many situations.

In this paper, we start by analyzing the failing cases of text-to-SQL tools and make the key observation that \emph{even when text-to-SQL tools return a wrong query, the structure of the query is often correct}. 
We use this insight to repair these failing cases with a mutation-based approach. When the text-to-SQL fails to synthesize a correct query, we ask the user for a complementary input-output example that will be used by our repair tool. We mutate the failing query until the output of the repaired query matches the example provided by the user. Our experimental evaluation shows that we can repair a large number of failing queries for two text-to-SQL tools, \ratsql~\cite{DBLP:conf/acl/WangSLPR20}, and \smbop~\cite{DBLP:conf/naacl/RubinB21}. Moreover, our approach also outperforms \patsql~\cite{patsql}, a state-of-the-art programming-by-example (PBE) synthesizer for SQL that uses examples, showing that we can successfully leverage the failing queries from text-to-SQL tools.
We summarize our contributions below.
\begin{enumerate}
    \item Key observation through an experimental analysis that when text-to-SQL tools return an incorrect query, it is often the case that the structure of the query is correct.
    \item Using our observation, we propose a mutation-based approach to repair queries agnostic to the text-to-SQL tool. %being used.
    \item We evaluate our approach on the Spider dataset\footnotemark[1]~and show that it can repair a large number of queries.
    \item We show that our approach outperforms and is complementary to PBE SQL synthesizers and improves the accuracy of text-to-SQL tools from 70\% to over 90\%.
\end{enumerate}

%\newpage
\section{Motivation}
\label{sec:motivation}

Recently, there has been a plethora of approaches 
using natural language processing to automatically synthesize SQL queries from text~\cite{DBLP:conf/acl/WangSLPR20,lin2020bridging,DBLP:conf/naacl/RubinB21,DBLP:conf/emnlp/ScholakSB21,DBLP:conf/emnlp/GanCXPWDZ21}. These approaches encode database relations using text-to-SQL encoders and have shown to have more than 70\% accuracy on public datasets such as Spider~\cite{DBLP:conf/emnlp/YuZYYWLMLYRZR18}. Spider is a large, complex, and cross-domain semantic parsing and text-to-SQL dataset. 
Even though text-to-SQL tools have shown remarkable performance, they do not always produce correct solutions. The goal of this work is to repair incorrect outputs into correct solutions. To motivate our repair technique, we first attempt to answer the following question:

\begin{mdframed}[style=mpdframe,frametitle=,innerleftmargin=5pt,innerrightmargin=5pt]
When text-to-SQL tools fail to find a correct query, how close is it to a solution?
\end{mdframed}

To answer the motivating question, we use the popular text-to-SQL tool  \ratsql~\cite{DBLP:conf/acl/WangSLPR20}, which uses relation-aware self-attention to map schema entities to SQL question words. 
We ran \ratsql
on 6358 queries from the Spider training dataset, and analyzed what happens when it fails to find a correct query. The Spider dataset contains a possible solution for each query, which we denote by \emph{gold query}. \ratsql returned a query\footnote{Even though \ratsql can return more than one query for a given natural language description, for this analysis we will considered the first query returned by \ratsql.} whose outputs do not match the gold query on 1197 of the 6358 (19\%) instances. 

We compare each of the 1197 faulty queries to its corresponding gold query as a \emph{string} and as a \emph{query structure} to determine how close the query is to a correct solution.

\subsection{String Analysis}

To determine whether the incorrect queries are significantly different from the correct version, we start by analyzing the \ratsql queries as strings with the gold query strings. We first normalize both strings by removing the capitalization and spaces. We then remove the table references for each column constant (e.g., ``airline.names" becomes ``names"). Finally, we calculate the edit distance \cite{ristad1998learning} between each incorrect query and its ground truth. 

\begin{figure}[!t]
	\centering
	\includegraphics[scale=0.95]{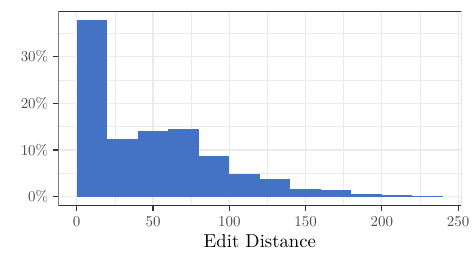}
	\vspace{-3mm}
	\caption{Distribution of \ratsql edit distances}
	\label{rat_edistance}
\end{figure}

Figure \ref{rat_edistance} shows the distribution of edit distances between the \ratsql query and the gold query. We observe that 36\% of queries have an edit distance larger than 60, suggesting that some queries are not close to the ground truth. However, we also observe that 38\% of \ratsql incorrect queries have edit distance below 20, with 18\% of the queries having an edit distance below 5. Our results suggest that \emph{when \ratsql fails, many queries are close to the gold query}. 

Even though string analysis gives us an intuition about the the failing cases, it is imprecise. To further strengthen our hypothesis, we also analyzed the structure of the failing queries.

\subsection{Structure Analysis}
To analyze the structure of the queries, we start by defining the \emph{query structure} of a SQL query in Definition~\ref{query_struct}. 

%To determine how many of the incorrect queries are structurally identical to the correct query, we parse each query into its query structure according to Definition \ref{query_struct}.
%

\begin{definition}[Query Structure]
Given a simple SQL query of the form:
\begin{minted}{sql}
SELECT A1, A2, ..., An
FROM R1, R2, .., Rm
WHERE P
\end{minted}

Then its corresponding query structure corresponds to:
\begin{minted}{sql}
SELECT #1, #2, ..., #n
FROM #(n+1), #(n+2), ..., #(n+m)
WHERE #(n+m+1)
\end{minted}

The SQL keywords 
(\texttt{SELECT}, \texttt{FROM}, and \texttt{WHERE}) are kept while the query constants (e.g., column names in a schema) are abstracted away and replaced with a placeholder \#i, where i is a unique integer that represents the placeholder. Note that the definition of query structure can be generalized for any SQL query, and it is not restricted to simple queries. 
\label{query_struct}
\end{definition}

To determine how many of the incorrect queries are structurally identical to the correct query, we parse each query into its corresponding query structure.
We observe that 790 out of 1197 (66\%) of the NLP queries have the same query structure as the correct query.

\begin{mdframed}[style=mpdframe,frametitle=,innerleftmargin=5pt,innerrightmargin=5pt]
Based on our edit distance results, we observe that
around half of the incorrect RAT-SQL queries have an edit distance less than 40 than the correct gold queries. We also observe that 790 out of 1197 (66\%) of the incorrect RAT-SQL queries have the same query structure as the correct query. Therefore, when text-to-SQL tools fail, they return a query that is close to the correct one.
Based on our findings, we propose building a mutation-based tool that automatically enumerates possible patches to repair the incorrect text-to-SQL queries.
\end{mdframed}

\section{Repairing Text-to-SQL Queries}
\label{sec:approach}

\begin{table*}[!t]
  \centering
  %\small
  \caption{Example of mutation types}
  \label{tab:mutation_type}
  \begin{tabular}{llll}
    \toprule
    Type & SQL Keywords & Mutation type & Example \\
    \midrule
    1 & SELECT, GROUPBY, ORDERBY & column/aggregate(column) & SELECT Students \\
    2 & FROM & table & FROM Classes \\
    3 & JOIN & table, column & JOIN Activities ON Students \\
    4 & WHERE, EXCEPT, HAVING & column, aggregate, constant & WHERE students LIKE `\%con\%' \\
    5 & LIMIT & constant & LIMIT 5 \\
    \bottomrule
  \end{tabular}%
  %\vspace{-2mm}
\end{table*}{}

\begin{figure*}[ht]
	\centering
	\includegraphics[width=\textwidth]{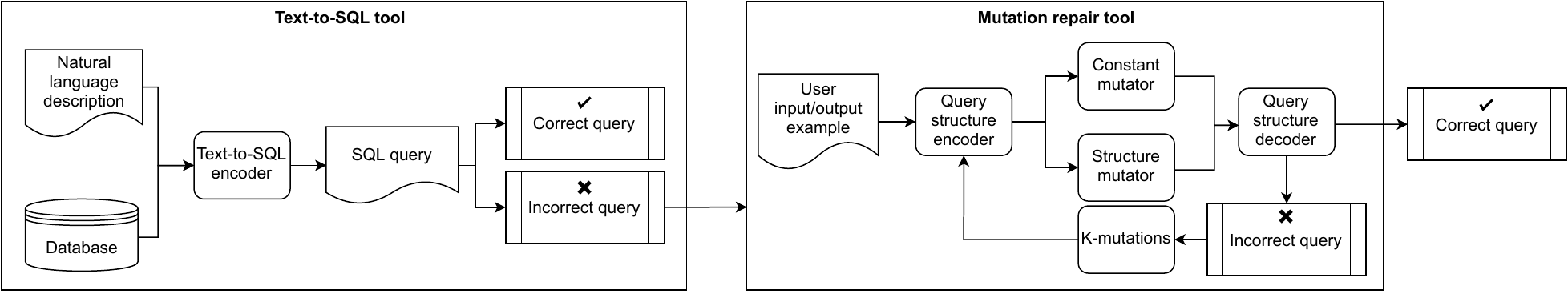} 
        \vspace{-5mm}
	\caption{Overview of our approach to repair text-to-SQL queries}
	%\caption{Overview of mutation repair's architecture}
	\label{architecture}
\end{figure*}

\begin{figure*}[ht]
	\centering
 \includegraphics[scale=0.72]{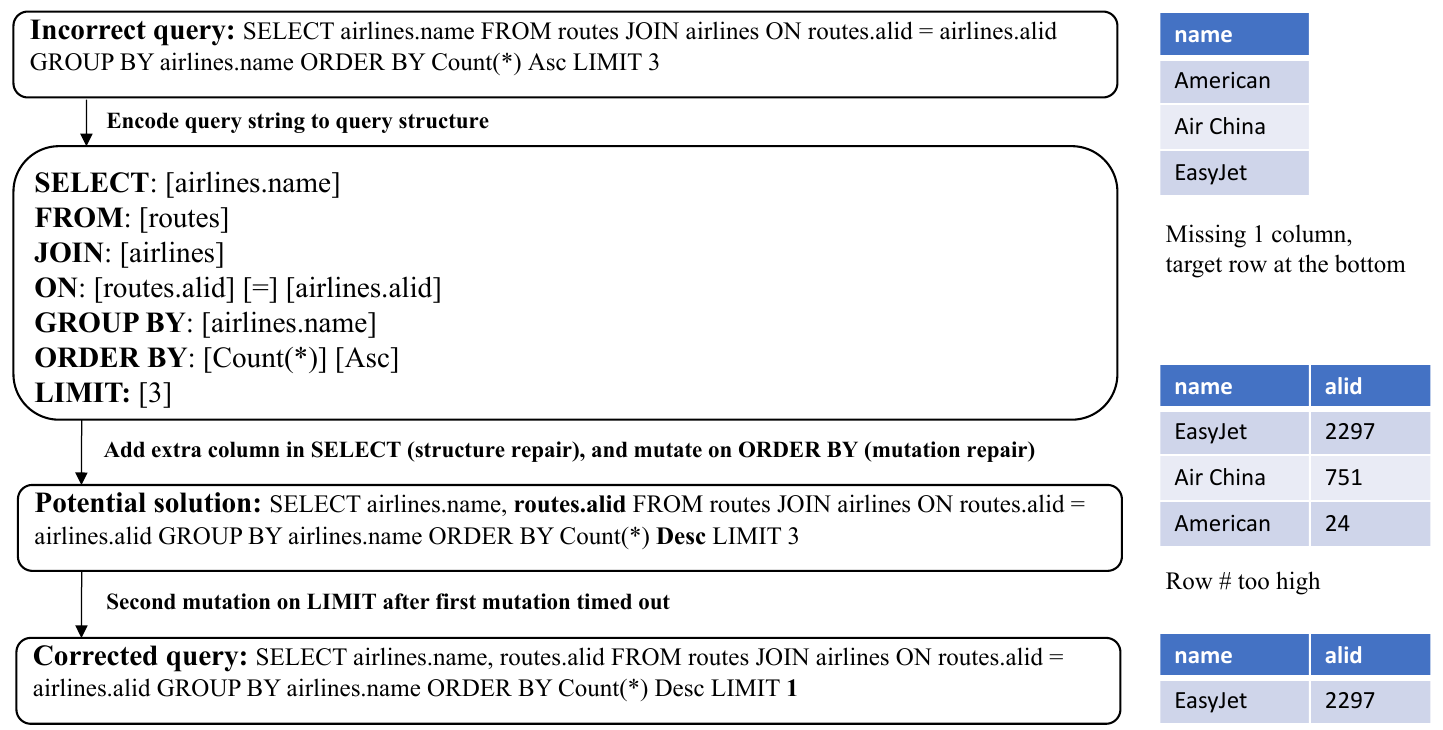} 
	\caption{Example repair of an incorrect RAT-SQL query}
	\label{mutation_example}
\end{figure*}

Figure \ref{architecture} shows an overview of our approach. First, we run a text-to-SQL tool that receives a natural language description in English and a database as input, and outputs a SQL query that captures the user intent. The user can then run the SQL query returned by the text-to-SQL tool and observe if the output is the expected one. If this is not the case, then the user provides additional information in the form of a small input-output example using tables that match the schema of the database. Synthesis of SQL using examples is also a common approach to solve this problem, and there are efficient synthesizers for this task~\cite{scythe,orvalhoVLDB20,patsql}. In our evaluation (\S\ref{sec:evaluation}), we show that our mutation-based approach can leverage the failed query from the text-to-SQL tool and outperform state-of-the-art SQL synthesizers based on examples. Note that our approach is orthogonal to the text-to-SQL tool used and can use any of the many available tools.

Our approach considers the incorrect SQL query from the text-to-SQL tool and starts by extracting its query structure. We identify each constant's type using the SQL keywords described in Table~\ref{tab:mutation_type}. We then mutate constants (``Constant mutator'') based on the keyword type (e.g., we mutate the table name for keyword \texttt{FROM}). For cases where a change in constants is not enough to repair the query, we perform a structural change (``Structure mutator'') in addition to the mutations of constants. If a single mutation is insufficient to repair the SQL query, our approach increases the number of mutations (``K-mutations'') and repeats this process.

\subsection{Single Mutation Repair}

Given a query structure, our mutation strategy is to select a placeholder $\#i$ and enumerate over all possible mutations that are compatible with the type of that placeholder. Since the search space for single mutations is relatively small, we can try this approach for all placeholders of a query structure and exhaust the space of single mutations. Consider the example in Figure~\ref{mutation_example} which corresponds to an incorrect SQL query returned by \ratsql for the example shown in \S\ref{sec:intro}. We first extract the query structure with placeholders represented in brackets. The query can be mutated by selecting a placeholder and changing that constant. For instance, we can select the placeholder that contains the constant `3' after the keyword \texttt{LIMIT} and change it to another constant value, e.g., `1'. Note that, for this particular example, we require a structural change since the keyword \texttt{SELECT} requires two columns in the correct solution and more than one mutation since we need to change the constants `Asc' to `Desc' and `3' to `1'.

\paragraph{Structural Repair.} \label{sec:struct} As observed in \S\ref{sec:motivation} and demonstrated in Figure~\ref{mutation_example}, there are some queries returned by text-to-SQL tools that do not match the desired query structure. To support some of these cases, we consider both the original query structure and minor modifications of the query structure. Namely, we consider changes that may involve adding \texttt{EXCEPT}, \texttt{WHERE} or \texttt{SELECT} clauses as follows.

\begin{itemize}
\item \textbf{EXCEPT clauses.}
Some queries may be missing an \texttt{EXCEPT} clause. If the output table of the incorrect query has more rows than expected, then we modify the query structure to include an \texttt{EXCEPT} clause and enumerate through all constants from previous \texttt{WHERE} clauses.
\item \textbf{WHERE clauses.} For queries that have output tables different than the expected size, we can also modify the query structure to either include a \texttt{WHERE} clause or modify existing \texttt{WHERE} clauses by adding another condition (conjunction or disjunction) to the \texttt{WHERE} clause.
\item \textbf{SELECT clauses}.
For query outputs that have the wrong number of columns, we add or remove a column, and enumerate over all possible choices for new columns to find the correct solution. Figure \ref{mutation_example} shows an example of using \texttt{SELECT} structural changes.
\end{itemize}

\paragraph{Dependency Analysis.} Trying all possible mutations of a given type may lead to mutations that are not executable. For instance, if the placeholder is of type column, trying columns that do not belong to a particular table will lead to an execution error. 
To reduce the number of infeasible mutations, we use information from the SQL schema to perform dependency analysis on the following types of mutation targets. 

\begin{itemize}
\item \textbf{Table dependency.} We enumerate through table mutation types for queries with \texttt{JOIN} statements based on their available columns. Our repair data set only includes inner joins. Inner join statements on two tables that do not have any overlap columns would not produce any results. Therefore, we do not enumerate tables in \texttt{JOIN} statements that do not share at least one column with preceding \texttt{FROM} statements.
\item \textbf{Column dependency.} For each column mutation type, we enumerate only column names corresponding to the query's current table names. We update the list of possible column names at each iteration of table mutation. 
\item \textbf{Where dependency.} The mutation type for constants depends on column types, and only mutations of that type are considered. To further reduce the search space, we use the results of the previous outputs for mutation choices of integer constants. Specifically, we relax the filter condition after observing a mutated output table with fewer rows than the correct table. Similarly, we tighten the filter condition when a mutated output table has more rows than the correct table. For example, if the output table from mutating ``\texttt{WHERE} student\_count $<$ 10" has too few rows, then we eliminate the possible mutation choices with integer values smaller than 10.

\end{itemize}

\subsection{Multi Mutation Repair}

Since some queries may require multiple mutations to be repaired (see Figure~\ref{mutation_example}), we extend our single mutation repair to a multi mutation approach with the following procedure:
\begin{enumerate}
\item Run our single mutation repair procedure on the initial query until completion.
\item For queries where we could not find a single mutation that repaired the initial query, we store all incorrect mutated queries and rank them based on a similarity metric with the expected output. 
\item For each top-ranked query in step (2), we consider that as our new initial query and repeat this procedure by going to step (1).
\end{enumerate}

Even though this procedure can repair multiple locations, it is not scalable. We limit our repair procedure to find at most 2 mutations. 
However, even with this limited approach, our experimental results (see \S\ref{sec:evaluation}) show that multi mutation repair can improve our performance and open research directions to consider more scalable algorithms for multiple repairs.

\paragraph{Similarity metric.} We rank the incorrect queries by comparing the contents of the output table of the incorrect query with the example provided by the user. Consider two multi-sets $\mathcal M$ and $\mathcal U$ that correspond to the values in the output table of the mutated query ($\mathcal M$) and the values in the example provided by the user ($\mathcal U$). We consider the Jaccard similarity coefficient between these two multi-sets as follows: 

\[
J(\mathcal M, \mathcal U) = \frac{\vert\mathcal M \cap \mathcal U\vert}{\vert\mathcal M\vert + \vert\mathcal U\vert - \vert\mathcal M \cap \mathcal U\vert}
\]

This similarity metric returns a value between 0 and 1, where 1 implies tables have the same contents, and 0 means that the table contents are disjoint.

\begin{figure*}[t]
    \centering
    \includegraphics[scale=0.95]{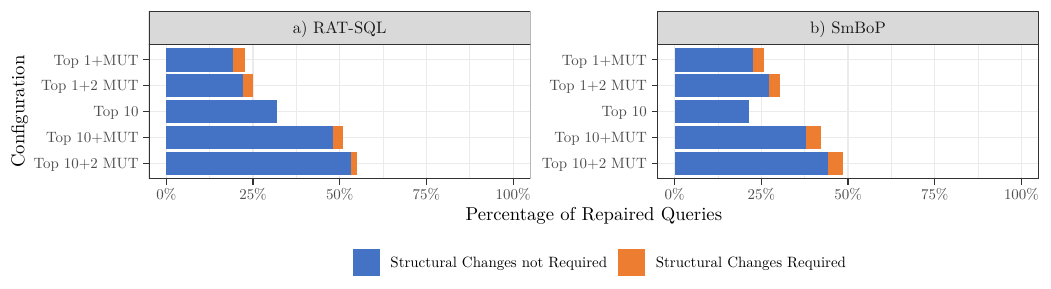}%
     %phantom subfigure needed for refs because the PDF contains both subplots 
     \begin{subfigure}{0pt}
     \phantomsubcaption
     \label{fig:ratsql}
     \end{subfigure}%
     \begin{subfigure}{0pt}
     \phantomsubcaption
     \label{fig:smbop}
     \end{subfigure}%
     \vspace{-3mm}
     \caption{Percentage of repaired queries for different tools and configurations.} \label{fig:results-structure}
\end{figure*}

\begin{figure*}[t]
\includegraphics[scale=0.95]{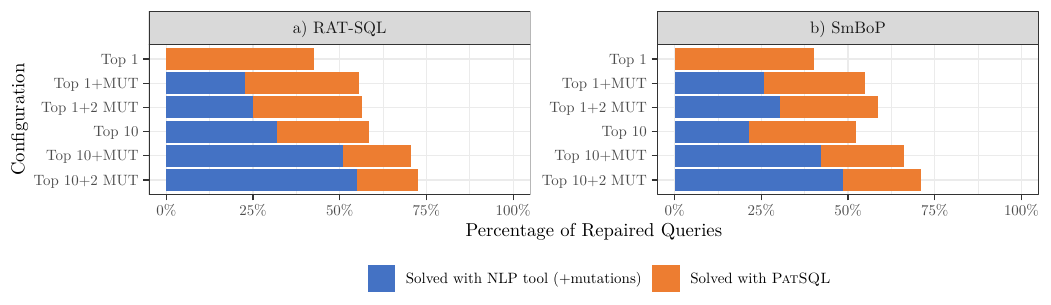}%
     %phantom subfigure needed for refs because the PDF contains both subplots 
     \begin{subfigure}{0pt}
     \phantomsubcaption
     \label{fig:ratsql-patsql}
     \end{subfigure}%
     \begin{subfigure}{0pt}
     \phantomsubcaption
     \label{fig:smbop-patsql}
     \end{subfigure}%
     \vspace{-3mm}
     \caption{Percentage of repaired queries using \patsql on top of our approach.}
     \label{fig:add-patsql}
\end{figure*}

\section{Evaluation}
\label{sec:evaluation}

In order to develop and test our repair tool, we use the Spider text-to-SQL dataset \cite{DBLP:conf/emnlp/YuZYYWLMLYRZR18}, consisting of 6358 instances in the training set and 1034 instances in the development set. In this evaluation, we focus on the development set. Since our work is centered around repairing queries produced by other tools, we consider two NLP to SQL synthesizers as a starting point: \ratsql \cite{DBLP:conf/acl/WangSLPR20} and \smbop \cite{DBLP:conf/naacl/RubinB21}. 
Nevertheless, our approach is agnostic to the text-to-SQL tool.

To test if a query satisfies the user intent, we generate an input-output example by running the gold query over the databases provided in the Spider dataset. Then we compare the output of the generated query, over that same input, with the output of the gold query.  All results were obtained using a dual socket Intel\textsuperscript{\tiny\textregistered} Xeon\textsuperscript{\tiny\textregistered} CPU E5-2630 with 64GB of RAM.

We note that the public \ratsql model provided on GitHub does not include the terminals in the generated queries. Hence, we created a terminal filler which uses three sources of knowledge: (1) the query without terminals, (2) the natural language question and (3) the database content. This constant filler, evaluated on the development set, has an accuracy of 91.4\%. This means that the constant extractor is able to correctly fill 457 out of 500 queries (from the 1034 queries in Spider development set, only 500 examples have terminals in its queries).

\subsection{Experimental Results}
\label{sec:results}

To evaluate the performance of our repair approach, we collect the top 10 queries generated by \ratsql and \smbop.
If we consider only the first query returned (i.e., the one deemed more likely to be correct), \ratsql is able to correctly solve 722/1034 instances, while \smbop correctly solves 783/1034. Moreover, if we consider the top 10 rated queries returned by the beam in each tool, \ratsql provides 822 correct queries, while \smbop provides 837. 

We consider as a baseline the number of instances solved when using just the first query returned by each of the tools. \autoref{fig:results-structure} shows the percentage of failed queries that we are able to repair when using different approaches.
In this context, 100\% means that all failed queries from the text-to-SQL tool were successfully repaired.

\autoref{fig:ratsql} shows the repair success rate on incorrect queries from \ratsql
when using one mutation (MUT) and two mutations (2 MUT). When only trying to repair the
first query (first two bars), we are able to repair 25\% of previously incorrect queries. 
We also observe that using the top 10 queries (Top 10), without any mutations, 
solves 32\% of the queries. When mutating the top 10, our approach is to attempt to repair each query, one by one and in order, until one of the queries is successfully repaired. Using these 10 queries as input to find a repair 
results in more than 50\% of repair success rate. Finally, note that using two mutations
allows the repair of 13 extra queries on top of the previous 50\%.

\autoref{fig:smbop} shows the repair success rate on incorrect queries from \smbop.
Overall, the repair success rate is similar to the ones in \ratsql.
Nevertheless, we can observe that using the top 10 queries does not have the same
impact as in \ratsql. This is due to two factors: (1) \smbop has an higher success rate
than \ratsql, failing only on harder instances, and (2) in general the 10 queries 
returned by \smbop are less diverse than the ones returned by \ratsql.

Figures~\ref{fig:ratsql} and~\ref{fig:smbop} also show the impact of repairing the query structure (\S\ref{sec:struct}). We observe that more instances require structural changes for \smbop than for \ratsql. Again, this results from less diversity in \smbop's query beam than in \ratsql's beam, i.e., it is more common in \ratsql that one of the top 10 queries has the correct structure.

\subsection{Improving results using \patsql}
\label{sec:improved-results}

We also leverage the input-output examples by running a programming 
by example SQL synthesizer. In this case, the SQL synthesizer is executed after our
mutation-based repair tool to find a query that satisfies the example provided by the user.
We choose \patsql~\cite{patsql} since it is one of the best performing SQL synthesizers.

Figure~\ref{fig:add-patsql} shows the repair success rate when executing \patsql
after our mutation-based repair tool for both \ratsql and \smbop.
The first bar (Top 1) corresponds to replacing our repair tool with \patsql.
However, the success rate of \patsql is smaller than our mutation-based repair tool.
For example, in repairing the \ratsql queries, \patsql has a success rate of
43\% while our tool repairs more than 50\% of the queries.
Nevertheless, by extending our mutation-based tool with \patsql, more than
70\% of failed queries can now be repaired. These improvements indicate
that \patsql is able to find correct queries that are structurally different 
from the ones returned by either \ratsql or \smbop.

\begin{figure}[t]
    \centering
    \includegraphics[scale=0.95]{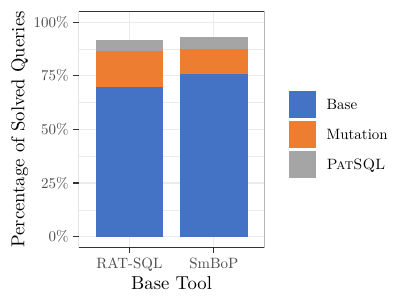}
    \vspace{-3mm}
    \caption{Overall results for \ratsql and \smbop.}
    \vspace{-3mm}
    \label{fig:results-overview}
\end{figure}

Finally, \autoref{fig:results-overview} shows the overall percentage of total queries 
solved using \ratsql and \smbop as a starting point. The `Base' part of the bar corresponds 
to using just the first query returned by the text-to-SQL tools, while the `Mutation' and  `\patsql' parts shows the contributions of our mutation-based approach and \patsql for the 
final result. 

\section{Related Work}
\label{sec:related}

\paragraph{Text-to-SQL}
Research in Natural Language (NL) to SQL semantic parsing has seen an increase in interest, in particular since the release of the Spider dataset \cite{DBLP:conf/emnlp/YuZYYWLMLYRZR18}.
Currently, most models designed for this task follow the encoder-decoder framework \cite{DBLP:conf/acl/WangSLPR20,lin2020bridging,DBLP:conf/naacl/RubinB21} using a neural language model  %(usually BERT or RoBERTa) 
and a regressive decoding strategy. Other improvements on this topic include pre-training strategies~\cite{yu2020grappa}.
% such as GraPPa

 \ratsql~\cite{DBLP:conf/acl/WangSLPR20} tackles this problem with schema-encoding and schema-linking mechanisms that convert the schema and the NL question into a graph. This is done by considering each word token as a vertex and its relations as the edges while further contextualizing them using relation-aware attention through transformer layers. 

\smbop~\cite{DBLP:conf/naacl/RubinB21} also uses the same encoding process as \ratsql, with the main distinction between the two being the decoding mechanism. \smbop decoder is semi-autoregressive, building the abstract syntax tree for the generated query in bottom-up style while \ratsql achieves this using a top-down autoregressive approach. 

\paragraph{Programming-by-Example}
Programming-by-Example SQL Synthesizers use one or more input-output examples to specify the behavior of the desired query. In recent years there has been a great number of tools released for this purpose, such as Scythe \cite{scythe}, SQUARES \cite{orvalhoVLDB20} and \patsql \cite{patsql}. In this work we use \patsql as a complementary approach since it performed the best out of the tested tools.

\patsql uses sketch-based enumeration in order to find a query that satisfies the input-output examples. This method works by first generating a skeleton of a query. Then, this skeleton is completed using the information from the tables and constants given by the user. \patsql introduces an improved completion step that is able to prune large parts of the program space and efficiently find solutions.

\paragraph{Program Repair}
The goal of program repair is to automatically repair a given buggy program. 
For that, it is necessary to locate the bug \cite{wong2013dstar,abreu2006evaluation} and 
create a \emph{patch} that repairs it. 
Mutation-based patch creation is common for program repair. For instance, GenProg \cite{le2011genprog} uses a search and mutation-based approach for repair of C programs.

Recent work uses program repair for SQL queries. Guo \textit{et al.} \cite{guo2018automatically} used fault localization and a decision tree (DT) algorithm for repairing \texttt{JOIN} and \texttt{WHERE} clauses. Presler-Marshall \textit{et al.} \cite{presler2021sqlrepair} used a sythesis based repair techniques on student authored SQL queries. Presler-Marshall \textit{et al.} found that SQL repair tools may be useful in an educational context. Our work is the first to use a mutation-based repair on all SQL clause types in the context of text-to-SQL queries.

\section{Conclusions}
\label{sec:conc}

SQL queries are challenging to write and require domain expertise. Recent text-to-SQL tools, such as \ratsql and \smbop, can return queries based on the user's intent written in natural language. However, even state-of-the-art text-to-SQL tools have an accuracy of just over 70\%. We observed that incorrect text-to-SQL queries are often structurally similar to the correct solution through experimental analysis. This work introduces a mutation-based repair tool that encodes incorrect queries into query structures and enumerates possible patches to produce a correct final query. 
Experimental results show that by leveraging a complementary input-output example provided by the user, our mutation-based approach can
improve the performance of text-to-SQL tools to 86\%.
Enhanced with a SQL synthesizer, the success rate of text-to-SQL tools
can be higher than 90\%.

\section*{Acknowledgments}
This work was partially supported by the US National Science Foundation (NSF) award CCF-1762363, and by
ANI 045917 award funded by FEDER and Portuguese Foundation for Science and Technology (FCT).

\bibliographystyle{named}
\bibliography{references}

\begin{thebibliography}{}

\bibitem[\protect\citeauthoryear{Abreu \bgroup \em et al.\egroup }{2006}]{abreu2006evaluation}
Rui Abreu, Peter Zoeteweij, and Arjan~JC Van~Gemund.
\newblock An evaluation of similarity coefficients for software fault localization.
\newblock In {\em Proc. Pacific Rim International Symposium on Dependable Computing}, pages 39--46. IEEE, 2006.

\bibitem[\protect\citeauthoryear{Cao \bgroup \em et al.\egroup }{2021}]{DBLP:conf/acl/CaoC0ZZ020}
Ruisheng Cao, Lu~Chen, Zhi Chen, Yanbin Zhao, Su~Zhu, and Kai Yu.
\newblock {LGESQL:} line graph enhanced text-to-sql model with mixed local and non-local relations.
\newblock In Chengqing Zong, Fei Xia, Wenjie Li, and Roberto Navigli, editors, {\em Proc. Annual Meeting of the Association for Computational Linguistics}, pages 2541--2555. Association for Computational Linguistics, 2021.

\bibitem[\protect\citeauthoryear{Gan \bgroup \em et al.\egroup }{2021}]{DBLP:conf/emnlp/GanCXPWDZ21}
Yujian Gan, Xinyun Chen, Jinxia Xie, Matthew Purver, John~R. Woodward, John~H. Drake, and Qiaofu Zhang.
\newblock Natural {SQL:} making {SQL} easier to infer from natural language specifications.
\newblock In {\em Proc. International Conference on Empirical Methods in Natural Language Processing}, pages 2030--2042. Association for Computational Linguistics, 2021.

\bibitem[\protect\citeauthoryear{Guo \bgroup \em et al.\egroup }{2018}]{guo2018automatically}
Yun Guo, Nan Li, Jeff Offutt, and Amihai Motro.
\newblock Automatically repairing sql faults.
\newblock In {\em Proc. IEEE International Conference on Software Quality, Reliability and Security}, pages 500--511. IEEE, 2018.

\bibitem[\protect\citeauthoryear{Le~Goues \bgroup \em et al.\egroup }{2011}]{le2011genprog}
Claire Le~Goues, ThanhVu Nguyen, Stephanie Forrest, and Westley Weimer.
\newblock Genprog: A generic method for automatic software repair.
\newblock {\em Ieee transactions on software engineering}, 38(1):54--72, 2011.

\bibitem[\protect\citeauthoryear{Lin \bgroup \em et al.\egroup }{2020}]{lin2020bridging}
Xi~Victoria Lin, Richard Socher, and Caiming Xiong.
\newblock {Bridging Textual and Tabular Data for Cross-Domain Text-to-SQL Semantic Parsing}.
\newblock In {\em Proc. International Conference on Empirical Methods in Natural Language Processing}, pages 4870--4888. Association for Computational Linguistics, 2020.

\bibitem[\protect\citeauthoryear{Orvalho \bgroup \em et al.\egroup }{2020}]{orvalhoVLDB20}
Pedro Orvalho, Miguel {Terra-Neves}, Miguel Ventura, Ruben Martins, and Vasco Manquinho.
\newblock {{SQUARES}}: A {{SQL}} synthesizer using query reverse engineering.
\newblock {\em Proceedings of the VLDB Endowment}, 13(12):2853--2856, August 2020.

\bibitem[\protect\citeauthoryear{Presler-Marshall \bgroup \em et al.\egroup }{2021}]{presler2021sqlrepair}
Kai Presler-Marshall, Sarah Heckman, and Kathryn~T Stolee.
\newblock {SQLRepair: Identifying and Repairing Mistakes in Student-Authored SQL Queries}.
\newblock In {\em Proc. IEEE/ACM International Conference on Software Engineering: Software Engineering Education and Training}, pages 199--210. IEEE, 2021.

\bibitem[\protect\citeauthoryear{Ristad and Yianilos}{1998}]{ristad1998learning}
Eric~Sven Ristad and Peter~N Yianilos.
\newblock Learning string-edit distance.
\newblock {\em IEEE Transactions on Pattern Analysis and Machine Intelligence}, 20(5):522--532, 1998.

\bibitem[\protect\citeauthoryear{Rubin and Berant}{2021}]{DBLP:conf/naacl/RubinB21}
Ohad Rubin and Jonathan Berant.
\newblock Smbop: Semi-autoregressive bottom-up semantic parsing.
\newblock In {\em {NAACL-HLT}}, pages 311--324. Association for Computational Linguistics, 2021.

\bibitem[\protect\citeauthoryear{Scholak \bgroup \em et al.\egroup }{2021}]{DBLP:conf/emnlp/ScholakSB21}
Torsten Scholak, Nathan Schucher, and Dzmitry Bahdanau.
\newblock {PICARD:} parsing incrementally for constrained auto-regressive decoding from language models.
\newblock In {\em Proc. International Conference on Empirical Methods in Natural Language Processing}, pages 9895--9901. Association for Computational Linguistics, 2021.

\bibitem[\protect\citeauthoryear{Shi \bgroup \em et al.\egroup }{2021}]{DBLP:conf/aaai/ShiNWZLWSX21}
Peng Shi, Patrick Ng, Zhiguo Wang, Henghui Zhu, Alexander~Hanbo Li, Jun Wang, C{\'{\i}}cero~Nogueira dos Santos, and Bing Xiang.
\newblock Learning contextual representations for semantic parsing with generation-augmented pre-training.
\newblock In {\em Proc. AAAI Conference on Artificial Intelligence}, pages 13806--13814. {AAAI} Press, 2021.

\bibitem[\protect\citeauthoryear{Takenouchi \bgroup \em et al.\egroup }{2021}]{patsql}
Keita Takenouchi, Takashi Ishio, Joji Okada, and Yuji Sakata.
\newblock {PATSQL:} efficient synthesis of {SQL} queries from example tables with quick inference of projected columns.
\newblock {\em Proc. {VLDB} Endow.}, 14(11):1937--1949, 2021.

\bibitem[\protect\citeauthoryear{Wang \bgroup \em et al.\egroup }{2017}]{scythe}
Chenglong Wang, Alvin Cheung, and Rastislav Bodik.
\newblock Synthesizing {{Highly Expressive SQL Queries}} from {{Input}}-output {{Examples}}.
\newblock In {\em Proc. Conference on Programming Language Design and Implementation}, pages 452--466, {New York, NY, USA}, 2017. {ACM}.

\bibitem[\protect\citeauthoryear{Wang \bgroup \em et al.\egroup }{2020}]{DBLP:conf/acl/WangSLPR20}
Bailin Wang, Richard Shin, Xiaodong Liu, Oleksandr Polozov, and Matthew Richardson.
\newblock {{RAT-SQL:} Relation-Aware Schema Encoding and Linking for Text-to-SQL Parsers}.
\newblock In {\em {ACL}}, pages 7567--7578. Association for Computational Linguistics, 2020.

\bibitem[\protect\citeauthoryear{Wong \bgroup \em et al.\egroup }{2013}]{wong2013dstar}
W~Eric Wong, Vidroha Debroy, Ruizhi Gao, and Yihao Li.
\newblock {The DStar method for effective software fault localization}.
\newblock {\em IEEE Transactions on Reliability}, 63(1):290--308, 2013.

\bibitem[\protect\citeauthoryear{Yaghmazadeh \bgroup \em et al.\egroup }{2017}]{sqlizer}
Navid Yaghmazadeh, Yuepeng Wang, Isil Dillig, and Thomas Dillig.
\newblock {{SQLizer}}: {{Query Synthesis}} from {{Natural Language}}.
\newblock {\em Proc. ACM Program. Lang.}, 1(OOPSLA):63:1--63:26, October 2017.

\bibitem[\protect\citeauthoryear{Yu \bgroup \em et al.\egroup }{2018}]{DBLP:conf/emnlp/YuZYYWLMLYRZR18}
Tao Yu, Rui Zhang, Kai Yang, Michihiro Yasunaga, Dongxu Wang, Zifan Li, James Ma, Irene Li, Qingning Yao, Shanelle Roman, Zilin Zhang, and Dragomir~R. Radev.
\newblock Spider: {A} large-scale human-labeled dataset for complex and cross-domain semantic parsing and text-to-sql task.
\newblock In {\em Proc. International Conference on Empirical Methods in Natural Language Processing}, pages 3911--3921. Association for Computational Linguistics, 2018.

\bibitem[\protect\citeauthoryear{Yu \bgroup \em et al.\egroup }{2021}]{yu2020grappa}
Tao Yu, Chien{-}Sheng Wu, Xi~Victoria Lin, Bailin Wang, Yi~Chern Tan, Xinyi Yang, Dragomir~R. Radev, Richard Socher, and Caiming Xiong.
\newblock Grappa: Grammar-augmented pre-training for table semantic parsing.
\newblock In {\em Proc. International Conference on Learning Representations}. OpenReview.net, 2021.

\end{thebibliography}

\end{document}